\documentclass[twocolumn,amsthm]{autartArXiv}
\pdfoutput=1

\usepackage{graphicx}

\usepackage[caption=false,font=normalsize,labelfont=sf,textfont=sf]{subfig}

\usepackage{amsmath}
\usepackage{amssymb}
\DeclareMathOperator*{\argmin}{arg\,min}
\DeclareMathOperator{\plim}{plim}

\newtheorem{notation}{Notation}
\newtheorem{definition}{Definition}
\newtheorem{assumption}{Assumption}
\newtheorem{theorem}{Theorem}
\newtheorem{lemma}{Lemma}
\newtheorem{remark}{Remark}

\usepackage[round,longnamesfirst]{natbib}

\usepackage[colorlinks=true,citecolor=blue]{hyperref}

\begin{document}

\hyphenation{
 		si-mu-la-tion
}

\begin{frontmatter}
\runtitle{Wiener system identification with GOBFs}

\title{Wiener system identification with generalized orthonormal basis functions\thanksref{footnoteinfo}}

\thanks[footnoteinfo]{This paper was not presented at any IFAC 
meeting. Corresponding author K.~Tiels. Tel. +32 2 6293665; 
fax +32 2 6292850.
}

\author{Koen Tiels}\ead{koen.tiels@vub.ac.be},
\author{Johan Schoukens}\ead{johan.schoukens@vub.ac.be}

\address{Vrije Universiteit Brussel, Department ELEC, Pleinlaan 2, B-1050 Brussels, Belgium}


\begin{abstract}
Many nonlinear systems can be described by a Wiener-Schetzen model. In this model, the linear dynamics are formulated in terms of orthonormal basis functions (OBFs). The nonlinearity is modeled by a multivariate polynomial. In general, an infinite number of OBFs is needed for an exact representation of the system. This paper considers the approximation of a Wiener system with finite-order infinite impulse response dynamics and a polynomial nonlinearity. We propose to use a limited number of generalized OBFs (GOBFs). The pole locations, needed to construct the GOBFs, are estimated via the best linear approximation of the system. The coefficients of the multivariate polynomial are determined with a linear regression. This paper provides a convergence analysis for the proposed identification scheme. It is shown that the estimated output converges in probability to the exact output.
Fast convergence rates, in the order $O_p({N_F}^{-n_{rep}/2})$, can be achieved, with $N_F$ the number of excited frequencies and $n_{rep}$ the number of repetitions of the GOBFs.
\end{abstract}

\end{frontmatter}

\section{Introduction}
\begin{figure}
	\centering
	\includegraphics[width = 0.25\textwidth]{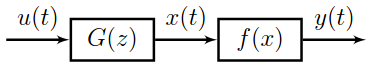}
	\caption{	A discrete-time SISO Wiener model
			($G$ is a linear dynamic system and $f$ is a nonlinear static system)
		 	\label{fig: Wiener model}}
\end{figure}

\begin{figure}
	\centering
	\includegraphics[width = 0.35\textwidth]{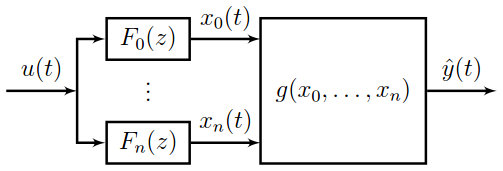}
	\caption{	A Wiener-Schetzen model
			(\mbox{$F_0, \ldots, F_n$} are scalar OBFs
			and $g$ is a multivariate polynomial function)
		 	\label{fig: Wiener-Schetzen model}}
\end{figure}

Even if nonlinear distortions are often present, many systems can be approximated by a linear model. When the nonlinear distortions are too large, a nonlinear model is required.
One option is to use a block-oriented model~\citep{Billings1982}, which consists of interconnections of linear dynamic and nonlinear static systems.
One of the simplest block-oriented models is the Wiener model (see Fig.~\ref{fig: Wiener model}). This is the cascade of a linear dynamic and a nonlinear static system. Wiener models have been used before to model e.g.\ biological systems~\citep{Hunter1986}, a pH process~\citep{Kalafatis1995}, and a distillation column~\citep{Bloemen2001}.
Some methods have been proposed to identify Wiener models, see e.g.~\citet{Greblicki1994} for a nonparametric approach where the nonlinearity is assumed to be invertible, \citet{Hagenblad2008} for the maximum likelihood estimator (MLE), \citet{Pelckmans2011} for an approach built on the concept of model complexity control, and \citet{Giri2013} for a frequency domain identification method where memory nonlinearities are considered. More complex parallel Wiener systems are identified in~\citet{Westwick1996} using a subspace based method, and in~\citet{Schoukens2012} using a parametric approach that needs experiments at several input excitation levels.
Some more Wiener identification methods can be found in the book edited by Giri and Bai~\citep{Giri2010}.
This paper considers a Wiener-Schetzen model (a type of parallel Wiener model, see Fig.~\ref{fig: Wiener-Schetzen model}), but, without loss of generality, the focus is on the approximation of a single-branch Wiener system, to keep the notation simple.

The recent book~\citet{Giri2010} also provides some industrial relevant examples of Wiener, Hammerstein, and Wiener-Hammerstein models, that can be handled by the Wiener-Schetzen model structure due to its parallel nature.
In fact, a Wiener-Schetzen model can describe a large class of nonlinear systems arbitrarily well in mean-square sense (see \citet{Schetzen2006} for the theory and other practical examples). If the system has fading memory, 
then for bounded slew-limited inputs,
a uniform convergence is obtained~\citep{Boyd1985}.
In a Wiener-Schetzen model, the dynamics are described in terms of orthonormal basis functions (OBFs). The nonlinearity is described by a multivariate polynomial.
Though any complete set of OBFs can be chosen, the choice is important for the convergence rate. If the pole locations of the OBFs match the poles of the underlying linear dynamic system closely, this linear dynamic system can be described accurately with only a limited number of OBFs~\citep{Heuberger2005}. In the original ideas of Wiener~\citep{Wiener1958}, Laguerre OBFs were used. Laguerre OBFs are characterized by a real-valued pole, making them suitable for describing well-damped systems with dominant first-order dynamics.
For moderately damped systems with dominant second-order dynamics, using Kautz OBFs is more appropriate~\citep{daRosa2007,VandenHof1995}.
We choose generalized OBFs (GOBFs), since they can deal with multiple real and complex valued poles~\citep{Heuberger2005}.

This paper considers the approximation of a Wiener system with finite-order IIR (infinite impulse response) dynamics and a polynomial nonlinearity by a Wiener-Schetzen model that contains a limited number of GOBFs.
The system poles are first estimated using the best linear approximation (BLA)~\citep{Pintelon2012} of the system. Next, the GOBFs are constructed using these pole estimates. The coefficients of the multivariate polynomial are determined with a linear regression.
The approach can be applied to parallel Wiener systems as well. As the estimation is linear-in-the-parameters, the Wiener-Schetzen model is well suited to provide an initial guess for nonlinear optimization algorithms, and for modeling time-varying and parameter-varying systems. The analysis in this paper is a starting point to tackle these problems.

The contributions of this paper are:
\begin{itemize}
	\item the proposal of an identification method for Wiener systems with finite-order IIR dynamics
(the initial ideas were presented in~\citet{Tiels2011}),
	\item a convergence analysis for the proposed method.
\end{itemize}

The paper is organized as follows. The basic setup is described in Section~\ref{sec: Setup}. Section~\ref{sec: Identification procedure} presents the identification procedure and the convergence analysis. Section~\ref{sec: Discussion} discusses the sensitivity to output noise. The identification procedure is illustrated on two simulation examples in Section~\ref{sec: Illustration}. Finally, the conclusions are drawn in Section~\ref{sec: Conclusion}.

\section{Setup}
\label{sec: Setup}

This section first introduces some notation, and next defines the considered system class and the class of excitation signals. Afterwards, the Wiener-Schetzen model is discussed in more detail. A brief discussion of the BLA concludes this section.

\subsection{Notation}
\label{sec: notation}

This section defines the notations $\plim$, $O(\cdot)$, and $O_p(\cdot)$.

\begin{notation}[$\plim$, \citep{Pintelon2012}]
	The sequence $x(N)$, \mbox{$N = 1, 2, \ldots$} converges to $x$ in probability if, for every \mbox{$\epsilon, \delta > 0$} there exists an $N_0$ such that for every \mbox{$N > N_0$ :} \mbox{$P(\left| x(N) - x \right| \le \epsilon) > 1 - \delta$}. We write
\begin{equation}
\nonumber
	\begin{gathered}
		\plim_{N \rightarrow \infty} x(N) = x\\
		\Leftrightarrow
		\forall \epsilon > 0 : \lim_{N \rightarrow \infty} P( \left| x(N) - x \right| \le \epsilon) = 1
	\end{gathered}
\end{equation}
\end{notation}

\begin{notation}[$O(\cdot)$]
	The notation $h_1$ is an $O(N^{\alpha})$ indicates that for $N$ big enough, \mbox{$\lvert h_1(N) \rvert \le c {N}^{\alpha}$}, where $c$ is a strictly positive real number.
\end{notation}

\begin{notation}[$O_p(\cdot)$, \citep{vanderVaart1998}]
	The notation $h_2$ is an $O_p({N}^{\alpha})$ indicates that the sequence $h_2(N)$ is bounded in probability at the rate $N^{\alpha}$. More precisely, \mbox{$h_2(N) = h_3(N) N^{\alpha}$}, where $h_3(N)$ is a sequence that is bounded in probability.
\end{notation}

\subsection{The Wiener system}
\label{sec: system}

The data-generating system is assumed to be a single input single output (SISO) discrete-time Wiener system (see Fig.~\ref{fig: Wiener model}). This is the cascade of a linear time-invariant (LTI) system $G(z)$ and a static nonlinear system $f(x)$.

In this paper, $G(z)$ is restricted to be a stable, rational transfer function, and $f(x)$ to be a polynomial.

\begin{definition}
	The class $\mathcal{G}$ is the set of proper, finite-dimensional, rational transfer functions, that are analytic in \mbox{$|z| \ge 1$} and squared integrable on the unit circle.
\end{definition}

\begin{assumption}
\label{assumption: G is a rational TF}
	The LTI system \mbox{$G(z) \in \mathcal{G}$}.
\end{assumption}

\begin{assumption}
\label{assumption: known order of G}
	The order of \mbox{$G(z) \in \mathcal{G}$} is known.
\end{assumption}
The poles of $G(z)$ are denoted by $p_j$ (\mbox{$j = 1, \ldots, n_p$}).

\begin{assumption}
\label{assumption: f non-even}
	The function $f(x)$ is non-even around the operating point.
\end{assumption}

\begin{assumption}
\label{assumption: f polynomial of known degree}
	The function $f(x)$ is a polynomial of known degree $Q$:
	\begin{equation}
	\label{eq: f(x)}
		f(x) = \sum_{p = 0}^{Q} \gamma_p x^p
		\quad .
	\end{equation}
\end{assumption}
More general functions can be approximated arbitrarily well in mean-square sense by~\eqref{eq: f(x)} over any finite interval.

The input $u(t)$ and the output \mbox{$y(t) = f(x(t))$} are measured at time instants \mbox{$t = k T_s$} (\mbox{$k = 0, \ldots, N - 1$}).

\subsection{Random-phase multisine excitation}
\label{sec: excitation signal}

In this paper, random-phase multisine~\citep{Pintelon2012} excitations are considered.

\begin{definition}
\label{def: random-phase multisine}
	A signal $u(t)$ is a random-phase multisine if~\citep{Pintelon2012}
	\begin{equation}
	\label{eq: multisine}
			u(t) =  \sum_{k = -N_F}^{N_F} U_k e^{j 2\pi \frac{k f_{max}}{N_F} t}
			\quad ,
	\end{equation}
	with \mbox{$U_k = U_{- k}^* = |U_k|e^{j\phi_k}$}, \mbox{$f_{max} = \frac{N_F}{N T_s}$} the maximum frequency of the excitation signal, \mbox{$N_F \le \frac{N}{2}$} the number of frequency components, and the phases $\phi_k$ uniformly distributed in the interval $[0,2\pi[$.
\end{definition}
The amplitudes $|U_k|$ can be chosen by the user, and are normalized such that $u(t)$ has finite power as \mbox{$N_F \rightarrow \infty$} \citep{Pintelon2012}.

\begin{definition}
\label{def: excitation signal class}
	The class of excitation signals $\mathcal{E}$ is the set of random-phase multisines $u(t)$, having normalized amplitudes \mbox{$U_k = \frac{1}{\sqrt{N_F}} \check{U}\left(\frac{k f_{max}}{N_F}\right)$}, where \mbox{$\check{U}(\frac{\omega}{2 \pi}) \in \mathbb{R}^{+}$} is a uniformly bounded function \mbox{($\check{U}(\frac{\omega}{2 \pi}) \le M_U/\sqrt{2} < \infty$)} with a countable number of discontinuities, and \mbox{$U_k = 0$} if \mbox{$\left| k \right| > N_F$} or \mbox{$k = 0$}.
\end{definition}

\begin{assumption}
\label{assumption: excitation signal}
	The excitation signal \mbox{$u(t) \in \mathcal{E}$}.
\end{assumption}
For simplicity, the excitations in this paper are restricted to random-phase multisines. However, as shown in~\citet{Schoukens2009a}, the theory applies for a much wider class of Riemann-equivalent signals. In this case, these are the extended Gaussian signals, which among others include Gaussian noise.

\subsection{The Wiener-Schetzen model}
\label{sec: model structure}

The system is modeled with a Wiener-Schetzen model (Fig.~\ref{fig: Wiener-Schetzen model}), where we choose \mbox{$F_1(z), \ldots, F_n(z)$} to be GOBFs.

In~\citet{Heuberger2005}, it is shown how a set of poles gives rise to a set of OBFs
\begin{subequations}
\label{eq: GOBFs}
	\begin{equation}
		F_l(z) = \frac{\sqrt{1 - |\xi_l|^2}}{z - \xi_l} \prod_{i = 1}^{l - 1} \left[ \frac{1 - \xi_i^{*} z}{z - \xi_i} \right]
		\quad .
	\end{equation}
If the poles $\xi_l$ result from a periodic repetition of a finite set of poles, the GOBFs are obtained, with poles
	\begin{equation}
		\xi_{j + (k - 1) n_{\xi}} = \xi_j
		\qquad j = 1, \ldots, n_{\xi} ; \, k = 1, 2, \ldots
		\quad .
	\end{equation}
\end{subequations}
The GOBFs form an orthonormal basis for the set of (strictly proper) rational transfer functions in~$\mathcal{G}$~\citep{Heuberger2005}.
One extra basis function is introduced, namely \mbox{$F_0(z) = 1$}, to enable the estimation of a feed-through term and as such also to enable the estimation of static systems~\citep{Tiels2011}. This extra basis function is still orthogonal with respect to the other basis functions (see Appendix~\ref{app: Orthonormal basis}).

The LTI system $G(z)$ can thus be represented exactly as a series expansion in terms of the basis functions $F_l(z)$:
\begin{equation}
\label{eq: series expansion G}
	G(z) = \sum_{l = 0}^{\infty} \alpha_l F_l(z)
	\quad .
\end{equation}
Let \mbox{$\hat{G}(z, n_{rep}) = \sum_{l = 0}^{n} \alpha_l F_l(z)$} be a truncated series expansion,
with \mbox{$n = n_{rep} n_{\xi}$}, and $n_{rep}$ the number of repetitions of the finite set of poles \mbox{$\{\xi_1, \ldots, \xi_{n_{\xi}}\}$}.
Recall that $p_j$ (\mbox{$j = 1, \ldots, n_p$}) are the true poles of $G(z)$, and let
\begin{equation}
\label{eq: rho}
	\rho = \max_j \prod_{k = 1}^{n_{\xi}} \left| \frac{p_j - \xi_{k}}{1 - p_j \xi_k} \right|
	\quad .
\end{equation}
Then there exists a finite \mbox{$c_{GOBF} \in \mathbb{R}$}, such that for any \mbox{$\eta \in \mathbb{R}$}, \mbox{$0 \le \rho < \eta < 1$}~\citep{deVries1998,Heuberger1995}
\begin{equation}
\label{eq: bound approx G}
	\Vert G(z) - \hat{G}(z, n_{rep}) \Vert_{\infty} \le c_{GOBF} \frac{\eta^{n_{rep}}}{1 - \eta}
	\quad ,
\end{equation}
which shows that $G(z)$ can be well approximated with a small number of GOBFs if the poles $\xi_j$ are close to the true poles $p_j$.
The pole locations $p_j$ will be estimated by means of the BLA of the system.

\subsection{The best linear approximation}
\label{sec: BLA}

The BLA of a system is defined as the linear system that approximates the system's output best in mean-square sense~\citep{Pintelon2012}.
The BLA of the considered Wiener system is equal to~\citep{Schoukens1998}
\begin{equation}
\label{eq: BLA}
	G_{BLA}(e^{j \omega_k}) = c_{BLA} \, G(e^{j \omega_k}) + O({N_F}^{-1})
	\quad ,
\end{equation}
where the constant $c_{BLA}$ depends upon the odd nonlinearities in $f(x)$ and the power spectrum of the input signal \mbox{$u(t) \in \mathcal{E}$}.
A similar result for Gaussian noise excitations results from Bussgang's theorem~\citep{Bussgang1952}.

\begin{remark}
	Under Assumption~\ref{assumption: f non-even}, $c_{BLA}$ is non-zero.
\end{remark}

\section{Identification procedure (no output noise)}
\label{sec: Identification procedure}

This section formulates the identification procedure and provides a convergence analysis in the noise-free case. The influence of output noise is analyzed in Section~\ref{sec: Discussion}.

The basic idea is that the asymptotic BLA (\mbox{$N_F \rightarrow \infty$}) has the same poles as $G(z)$ (see~\eqref{eq: BLA}). The poles calculated from the estimated BLA are thus excellent candidates to be used in constructing the GOBFs. Since the BLA will be estimated from a finite data set ($N_F$ finite), the poles calculated from the estimated BLA will differ from the true poles. Extensions of the basis functions (\mbox{$n_{rep} > 1$}) will be used to compensate for these errors (see~\eqref{eq: bound approx G}).

The identification procedure can be summarized as:
\begin{enumerate}
	\item Estimate the BLA and calculate its poles.
	\item Use these pole estimates to construct the GOBFs.
	\item Estimate the multivariate polynomial coefficients.

\end{enumerate}
These steps  are now formalized and the asymptotic behavior (\mbox{$N_F \rightarrow \infty$}) of the estimator is analyzed. First, the situation without disturbing noise is considered. The influence of disturbing noise is discussed in Section~\ref{sec: Discussion}.

\subsection{Identify the BLA and calculate its poles}
\label{sec: step 1a}

\subsubsection{Nonparametric and parametric BLA}

First, a nonparametric estimate of the BLA is calculated. Since the input is periodic, the BLA is estimated as
\mbox{$\hat{G}_{BLA}(e^{j \omega_k}) = \frac{Y(e^{j \omega_k})}{U(e^{j \omega_k})}$},
in which $Y(e^{j \omega_k})$ and $U(e^{j \omega_k})$ are the discrete Fourier transforms (DFTs) of the output and the input.
For random excitations, the classical frequency response estimates (division of cross-power and auto-power spectra) can be used~\citep{Pintelon2012}, or more advanced FRF measurement techniques can be used, like the local polynomial method~\citep{Pintelon2010}.

Next, a parametric model is identified using a weighted least-squares estimator~\citep{Schoukens1998}
\begin{subequations}
\label{eq: theta_hat}
	\begin{equation}
		\hat{\theta}(N_F) = \argmin_{\theta} K_{N_F}(\theta)
		\quad ,
	\end{equation}
where the cost function $K_{N_F}(\theta)$ is equal to
 	\begin{equation}
	\label{subeq: theta_hat}
		\frac{1}{N_F} \sum_{k = 1}^{N_F} 
				W(k) \left| \hat{G}_{BLA}(e^{j \omega_k}) - G_M(e^{j \omega_k},\theta) \right|^2
		\quad .
	\end{equation}
Here, \mbox{$W(k) \in \mathbb{R}^{+}$} is a deterministic, $\theta$\nobreakdash-\hspace{0pt}independent weighting sequence, and $G_M(e^{j \omega_k}, \theta)$ is a parametric transfer function model
	\begin{equation}
		\begin{gathered}
			G_M(e^{j \omega_k},\theta)	= \frac	{\sum_{l = 0}^{n_b} b_l e^{-j \omega_k l}}
									{\sum_{l = 0}^{n_a} a_l e^{-j \omega_k l}}	
							= \frac 	{B_{\theta}(e^{j \omega_k})}
									{A_{\theta}(e^{j \omega_k})}
			\quad , \\
			\theta = \begin{bmatrix}a_0 & \cdots & a_{n_a} & b_0 & \cdots & b_{n_b}\end{bmatrix}^{T}
			\quad ,
		\end{gathered}
	\end{equation}
with the constraint \mbox{$\Vert \theta \Vert_2 = 1$} to obtain a unique parameterization. Under Assumption~\ref{assumption: known order of G}, we put \mbox{$n_a = n_p$}.
\end{subequations}

\subsubsection{Pole estimates}

Eventually, the poles $\hat{p}_j$ (\mbox{$j = 1, \ldots, n_p$}) of the parametric model $G_M(e^{j \omega_k},\hat{\theta})$ are calculated.
Before we derive a bound on \mbox{$\Delta p_j := \hat{p}_j - p_j$} in Lemma~\ref{lemma: bound on delta_p}, a regularity condition on the parameter set $\theta$ is needed.
\begin{assumption}
\label{assumption: regularity condition theta}
	The parameter set $\theta$ is identifiable if the system is excited by \mbox{$u(t) \in \mathcal{E}$}.
\end{assumption}
The existence of a uniformly bounded convergent Volterra series~\citep{Schetzen2006,Schoukens1998} is needed as well (see Appendix~\ref{app: Volterra series} for more details).
\begin{assumption}
\label{assumption: convergent Volterra series}
	There exists a uniformly bounded Volterra series whose output converges in least-squares sense to the true system output for \mbox{$u(t) \in \mathcal{E}$}.
\end{assumption}

\begin{lemma}
\label{lemma: bound on delta_p}
Consider a discrete-time Wiener system, with an LTI system $G(z)$ and a static nonlinear system $f(x)$. Let $p_j$ (\mbox{$j = 1, \ldots, n_p$}) be the poles of $G(z)$ and $\hat{p}_j$ be the pole estimates, obtained using the weighted least-squares estimator~\eqref{eq: theta_hat}. Then under Assumptions~\mbox{\ref{assumption: G is a rational TF} -- \ref{assumption: f non-even}}, and \mbox{\ref{assumption: excitation signal} -- \ref{assumption: convergent Volterra series}}, \mbox{$\Delta p_j := \hat{p}_j - p_j$} is an \mbox{$O_p({N_F}^{-1/2})$}.
\end{lemma}
\begin{pf}
Let $\tilde{\theta}$ be the ``true'' model parameters, such that
\begin{equation}
\label{eq: theta_BLA}
	\begin{aligned}
		G_M(e^{j \omega_k}, \tilde{\theta}) 	& = c_{BLA} G(e^{j \omega_k})\\
								& = \frac 	{B_{\tilde{\theta}}(e^{j \omega_k})}
										{A_{\tilde{\theta}}(e^{j \omega_k})}
	\end{aligned}
	\quad \forall \omega_k
	\quad ,
\end{equation}
with $A_{\tilde{\theta}}$ and $B_{\tilde{\theta}}$ polynomials of degree $n_p$.
The roots of $A_{\tilde{\theta}}(e^{j \omega_k})$ are equal to the true poles $p_j$ (\mbox{$j = 1, \ldots, n_p$}).
If these poles are all distinct, the first-order Taylor expansion of $A_{\tilde{\theta}}(e^{j \omega_k})$ results in~\citep{Guillaume1989}
\begin{subequations}
\label{eq: sensitivity poles}
	\begin{equation}
		\Delta p_j \approx - \sum_{l = 0}^{n_a} \frac{{p_j}^l}{A_{\tilde{\theta}}^{'}(p_j)} \Delta a_l
		\quad ,
	\end{equation}
where $A_{\tilde{\theta}}^{'}(p_j) \ne 0$, and where $\Delta a_l$ follows from
	\begin{equation}
		\hat{\theta} - \tilde{\theta} = 
			\begin{bmatrix} 
				\Delta a_0 & \cdots & \Delta a_{n_a} & \Delta b_0 & \cdots & \Delta b_{n_b}
			\end{bmatrix}^{T}
		\quad .
	\end{equation}
\end{subequations}
Under Assumptions~\mbox{\ref{assumption: excitation signal} -- \ref{assumption: convergent Volterra series}}, it is shown in~\citet{Schoukens1998} that
\mbox{$\plim_{N_F \rightarrow \infty} \left( \hat{\theta}(N_F) - \tilde{\theta} \right) = 0$}.
For the considered output disturbances (no noise in this section, filtered white noise in Section~\ref{sec: Discussion}), the least-squares estimator~\eqref{eq: theta_hat} is a MLE~\citep{Pintelon1994}.
From the properties of the MLE, it follows that~\citep{Pintelon2012}
\begin{equation}
\label{eq: bound hat_theta}
	\hat{\theta} = \tilde{\theta} + O_p({N_F}^{-1/2})
	\quad .
\end{equation}
Then from~\eqref{eq: sensitivity poles} and~\eqref{eq: bound hat_theta}, it follows that
\begin{equation}
\label{eq: bound Delta_p_j}
	\Delta p_j = O_p({N_F}^{-1/2})
	\quad .
\end{equation}
This concludes the proof of Lemma~\ref{lemma: bound on delta_p}. \qed
\end{pf}
Lemma~\ref{lemma: bound on delta_p} shows that good pole estimates are obtained.
\begin{remark}
Though no external noise is considered in this section, the probability limits in this paper are w.r.t. the random phase realizations of the excitation signal.
\end{remark}

\subsection{Construct the GOBFs}
\label{sec: step 2}

Next, the GOBFs are constructed with these pole estimates (see~\eqref{eq: GOBFs}, with \mbox{$\xi_j = \hat{p}_j$}), and the intermediate signals \mbox{$x_l(t) = F_l(z) u(t)$} (\mbox{$l = 0, \ldots, n$}) (see Fig.~\ref{fig: Wiener-Schetzen model}) are calculated. The following lemma shows that the true intermediate signal $x(t)$ can be approximated arbitrarily well by a linear combination of the calculated signals $x_l(t)$.%
{\sloppy
\begin{lemma}
\label{lemma: bound on delta_x}
Consider the situation of Lemma~\ref{lemma: bound on delta_p}. Let \mbox{$F_0(z) = 1$}, and let $F_l(z)$ (\mbox{$l = 1, 2, \ldots$}) be GOBFs, constructed from the finite set of poles \mbox{$\{ \hat{p}_1, \ldots, \hat{p}_{n_p} \}$}. Let \mbox{$x(t) = G(z) u(t)$} and \mbox{$x_l(t) = F_l(z) u(t)$}. Let \mbox{$G(z) = \sum_{l = 0}^{\infty} \alpha_l F_l(z)$}, and denote \mbox{$n = n_{rep} n_p$}. Then under Assumptions~\mbox{\ref{assumption: G is a rational TF} -- \ref{assumption: f non-even}}, and \mbox{\ref{assumption: excitation signal} -- \ref{assumption: convergent Volterra series}}, \mbox{$\Delta x(t) := x(t) - \sum_{l = 0}^{n} \alpha_l x_l(t)$} is an \mbox{$O_p({N_F}^{-n_{rep}/2})$}.
\end{lemma}}
\begin{pf}
From~\eqref{eq: rho} and~\eqref{eq: bound Delta_p_j}, it follows that $\rho$ is an $O_p({N_F}^{-1/2})$. It then follows  from~\eqref{eq: bound approx G} that
\mbox{$G(z) - \sum_{l = 0}^{n} \alpha_l F_l(z) = O_p({N_F}^{-n_{rep}/2})$},
and thus \mbox{$\Delta x(t)$} is an \mbox{$O_p({N_F}^{-n_{rep}/2})$}. \qed
\end{pf}

\subsection{Estimate the multivariate polynomial coefficients}
\label{sec: step 3}
Finally, the coefficients of the multivariate polynomial \mbox{$g(x_0, \ldots, x_n)$} are estimated.
Let
\begin{subequations}
\label{eq: y_beta}
\begin{equation}
 	y_{\beta} = \beta_{DC} + \sum_{p = 1}^{Q} \left( 
			\sum_{i_1 = 0}^{n} \sum_{i_2 = i_1}^{n} \cdots \sum_{i_p = i_{p - 1}}^{n} 
			\beta_{i_1, \ldots, i_p} x_{i_1} \cdots x_{i_p} \right)
\end{equation}
be the output of \mbox{$g(x_0, \ldots, x_n)$}, where the coefficients of the polynomial are chosen to be
\begin{equation}
	\beta =
	\begin{bmatrix}
		\beta_{DC} & \beta_{0} & \cdots  & \beta_{i_1, \ldots, i_p} & \cdots  & \beta_{n, \ldots, n}
	\end{bmatrix}^{T}
	\quad .
\end{equation}
\end{subequations}
They are estimated using linear least-squares regression:
\begin{subequations}
\label{eq: beta_hat}
\begin{equation}
	\hat{\beta}	= \argmin_{\beta} \Vert y_{\beta} - y \Vert_2
			= \argmin_{\beta} \Vert \Psi \beta - y \Vert_2
	\quad ,
\end{equation}
where the regression matrix $\Psi$ is equal to
\begin{equation}
	\Psi =
	\begin{bmatrix}
		1				& \cdots 	& 1						\\
		x_0(0)				& \cdots 	& x_0(N - 1)					\\
		\vdots 				& \ddots 	& \vdots 					\\
		x_{i_1}(0) \cdots x_{i_p}(0)	& \cdots 	& x_{i_1}(N - 1) \cdots x_{i_p}(N - 1)	\\
		\vdots 				& \ddots 	& \vdots 					\\
		{x_n}^{Q}(0)			& \cdots 	& {x_n}^{Q}(N - 1)
	\end{bmatrix}^{T}
\end{equation}
\end{subequations}
We will now show that the estimated output \mbox{$\hat{y}(t) := y_{\hat{\beta}}$} converges in probability to $y(t)$ as \mbox{$N_F \rightarrow \infty$}.

\begin{theorem}
\label{theorem: convergence y_hat}
Consider the situation of Lemma~\ref{lemma: bound on delta_x}.
Let \mbox{$y(t) = f(x(t))$} and
\mbox{$\hat{y}(t) = y_{\hat{\beta}}(t)$}, 
where the coefficients $\hat{\beta}$ are obtained from the least-squares regression~\eqref{eq: beta_hat}. Then under Assumptions~\mbox{\ref{assumption: G is a rational TF} -- \ref{assumption: convergent Volterra series}}, \mbox{$\hat{y}(t) - y(t)$} is an \mbox{$O_p({N_F}^{-n_{rep}/2})$}.
\end{theorem}
\begin{pf}
The exact output
\begin{equation}
\label{eq: exact y}
	\begin{aligned}
		y(t)	& = \sum_{p = 0}^{Q} \gamma_p x^p(t)\\
			& = \sum_{p = 0}^{Q} \gamma_p \left( \sum_{l = 0}^{n} \alpha_l x_l(t) + \Delta x(t) \right)^p\\
			& = \left[ \sum_{p = 0}^{Q} \gamma_p \left(
				\sum_{l = 0}^{n} \alpha_l x_l(t) \right)^p \right] + \Delta y\\
			& = 
				\begin{aligned}[t]
					& \tilde{\beta}_{DC}\\
					& + \sum_{p = 1}^{Q} \left( 
						\sum_{i_1 = 0}^{n} \sum_{i_2 = i_1}^{n} \cdots \sum_{i_p = i_{p - 1}}^{n}
						\tilde{\beta}_{i_1, \ldots, i_p} x_{i_1} \cdots x_{i_p} \right)\\
					& + \Delta y
				\end{aligned}\\
			& = y_{\tilde{\beta}} + \Delta y
	\end{aligned}
\end{equation}
where $y_{\tilde{\beta}}$ is the output of a multivariate polynomial \mbox{$g(x_0, \ldots, x_n)$}, in which the coefficients $\tilde{\beta}$ follow from the true coefficients $\gamma_p$ of $f(x)$ and the true coefficients $\alpha_l$ of the series expansion of $G(z)$. The truncation of this series expansion is taken into account by the term \mbox{$\Delta y$}, which just as \mbox{$\Delta x$} is an $O_p({N_F}^{-n_{rep}/2})$.

Note that the coefficients $\tilde{\beta}$ can be obtained as the minimizers of the artificial least-squares problem
\begin{equation}
\label{eq: beta_tilde}
	\tilde{\beta} = \argmin_{\beta} \Vert \Psi \beta - y_{\tilde{\beta}} \Vert_2
	\quad .
\end{equation}
The estimated coefficients are equal to
\begin{equation}
\label{eq: beta_hat 2}
	\begin{aligned}
		\hat{\beta}	& = (\Psi^{T} \Psi)^{-1} \Psi^{T} y\\
				& = (\Psi^{T} \Psi)^{-1} \Psi^{T} (y_{\tilde{\beta}} + \Delta y)\\
				& = \tilde{\beta} + (\Psi^{T} \Psi)^{-1} \Psi^{T} \Delta y
	\end{aligned}
	\quad .
\end{equation}
We now show that \mbox{$(\Psi^{T} \Psi)^{-1} \Psi^{T} \Delta y$} is an \mbox{$O_p({N_F}^{-n_{rep}/2})$}.

Each element in $\Psi$ is an $O({N_F}^0)$, due to the normalization of the excitation signal (see Assumption~\ref{assumption: excitation signal}). Consequently, each element in the matrix \mbox{$\Psi^{T} \Psi$} is the sum of $N$ terms that are an $O({N_F}^0)$, so \mbox{$\Psi^{T} \Psi$} is an $O(N)$. The elements in the matrix \mbox{$(\Psi^{T} \Psi)^{-1}$} are thus an $O(N^{-1})$.

Each element in the vector \mbox{$\Delta y$} is an $O_p({N_F}^{-n_{rep}/2})$. Consequently, each element in the vector \mbox{$\Psi^{T} \Delta y$} is the sum of $N$ terms that are the product of an $O({N_F}^0)$ and an $O_p({N_F}^{-n_{rep}/2})$. The elements in the vector \mbox{$\Psi^{T} \Delta y$} are thus an \mbox{$O(N) O_p({N_F}^{-n_{rep}/2})$}.

As a consequence, \mbox{$\hat{\beta} = \tilde{\beta} + O_p({N_F}^{-n_{rep}/2})$}. And thus
\begin{equation}
	\hat{y}(t) - y(t)	= \Psi (\hat{\beta} - \tilde{\beta}) - \Delta y
				= O_p({N_F}^{-n_{rep}/2})
	\quad .
\end{equation}
This concludes the proof of Theorem~\ref{theorem: convergence y_hat}. \qed
\end{pf}
Theorem~\ref{theorem: convergence y_hat} shows that the estimated output converges in probability to the exact output with only a finite number of basis functions. The convergence rate increases if $n_{rep}$ is increased.
\begin{remark}
	The multivariate polynomial $g(x_0, \ldots, x_n)$ is implemented in terms of Hermite polynomials~\citep{Schetzen2006} to improve the numerical conditioning of the least-squares estimation in~\eqref{eq: beta_hat}~\citep{Tiels2011}. As this has no consequences for the result in Theorem~\ref{theorem: convergence y_hat}, ordinary polynomials are used throughout the paper to keep the notation simple.
\end{remark}

\section{Noise analysis}
\label{sec: Discussion}

In this section, the sensitivity of the identification procedure to output noise is discussed. Noise on the intermediate signal $x(t)$ is not considered here. In general, this would result in biased estimates of the nonlinearity. More involved estimators, e.g. the MLE, are needed to obtain an unbiased estimate~\citep{Hagenblad2008,Wills2013}.

In the case of filtered white output noise \mbox{$v(t) = H(q) e(t)$}, with $e(t)$ a sequence of independent random variables, independent of $u(t)$, with zero mean and variance $\lambda$, and with $H(q)$ a stable monic filter, the exact output \mbox{$y(t) = y_{\tilde{\beta}}(t) + \Delta y(t) + v(t)$}. The estimated coefficients are then equal to (cfr.~\eqref{eq: beta_hat 2})
\begin{equation}
	\hat{\beta} = \tilde{\beta} + (\Psi^{T} \Psi)^{-1} \Psi^{T} \Delta y + (\Psi^{T} \Psi)^{-1} \Psi^{T} v
	\quad .
\end{equation}
The columns of $\Psi$ are filtered versions of the known input signal $u(t)$, which was assumed independent of $v(t)$. It is thus clear that the noise $v(t)$ is uncorrelated with the columns of $\Psi$.
Consequently, each element in the vector \mbox{$\Psi^{T} v$} is the sum of $N$ uncorrelated terms that are the product of an $O({N_F}^0)$ and an $O(N^0)$. The elements in the vector \mbox{$\Psi^{T} v$} are thus an $O(N^{1/2})$, As a consequence, \mbox{$(\Psi^{T} \Psi)^{-1} \Psi^{T} v = O(N^{-1/2})$}.

The error on the estimated output due to the noise is thus independent of the number of repetitions $n_{rep}$. Increasing $n_{rep}$ allows to tune the model error such that it disappears in the noise floor.

\section{Illustration}
\label{sec: Illustration}

In this section, the approach is illustrated on two simulation examples. The first one considers the noise-free case, and illustrates the convergence rate predicted by Theorem~\ref{theorem: convergence y_hat}. The second one compares the proposed method to the so-called approximative prediction error method (PEM)~\citep{Hagenblad2008}, as implemented in the MATLAB system identification toolbox~\citep{Ljung2013}.

\subsection{Example~1: noise-free case}

Consider a SISO discrete-time Wiener system with
\begin{equation}
	G(z) = \frac{1 + 3 z^{-1} + 3 z^{-2} + z^{-3}}{1 - 2.1 z^{-1} + 1.9 z^{-2} - 0.7 z^{-3}}
	\quad ,
\end{equation}
and
\begin{equation}
	f(x) = x + 0.8 x^2 + 0.7 x^3
	\quad .
\end{equation}
The system is excited with a random-phase multisine (see~\eqref{eq: multisine}) with \mbox{$f_{max} = f_s/6$} and $f_s$ the sampling frequency; \mbox{$N_F = 170, 341, 682, 1365, 2730, 5461, 10922$}; and the amplitudes \mbox{$| U_k |$} chosen equal to each other and such that the rms value of $u(t)$ is equal to $1$.
The system is identified using the identification procedure described in Section~\ref{sec: Identification procedure}. No weighting is used to obtain a parametric estimate of the BLA, i.e. \mbox{$W(k) = 1$ in~\eqref{subeq: theta_hat}}.
A random-phase multisine with \mbox{$N_F = 10922$} is used for the validation. Fifty Monte Carlo simulations are performed, with each time a different realization of the random phases of the excitation signals.

\begin{figure}
	\centering
	\includegraphics[width=0.4\textwidth]{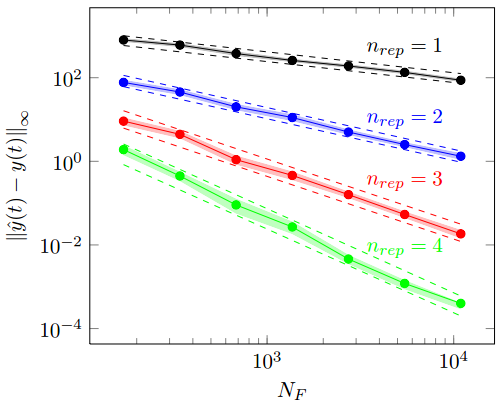}
	\caption{Average of $\left\Vert \hat{y}(t) - y(t) \right\Vert_\infty$ along the $50$ Monte Carlo simulations (full line) and its one standard deviation confidence interval (filled zone). The predicted convergence rate is indicated by the dashed lines, which are an $O({N_F}^{-n_{rep}/2})$.
\label{fig: Monte Carlo}}
\end{figure}

The results in Fig.~\ref{fig: Monte Carlo} show that the convergence rate of \mbox{$(\hat{y}(t) - y(t))$} agrees with what is predicted by Theorem~\ref{theorem: convergence y_hat}. The convergence rate increases with an increasing number of repetitions of the basis functions.
These results generalize to parallel Wiener systems as well.

\subsection{Example~2: noisy case with saturation nonlinearity}

The second example is inspired by the second example in~\citet{Hagenblad2008}. It is a discrete-time SISO Wiener system with a saturation nonlinearity. The system is given by
\begin{equation}
\label{eq: example 2}
	\begin{aligned}
		& x(t) + 0.3 x(t - 1) - 0.3 x(t - 2)					\\
		&	\quad = u(t) - 0.3 u(t - 1) + 0.3 u(t - 2)				\\
		& f(x(t)) = 
			\begin{cases}
				c_1	&	\text{for } x(t) \le c_1			\\
				x(t)	&	\text{for } c_1 < x(t) \le c_2		\\
				c_2	&	\text{for } c_2 < x(t)
			\end{cases}								\\
		& y(t) = f \left( x(t) \right) + e(t)
	\end{aligned}
	\quad ,
\end{equation}
where the input $u$ and the output noise $e$ are Gaussian, with zero mean, and with variances \mbox{$\lambda_u= 1$} and \mbox{$\lambda_e = 0.01$} respectively.
The coefficients $c_1$ and $c_2$ are equal to $-0.4$ and $0.2$, respectively.
Compared to the example in~\citet{Hagenblad2008}, no process noise was added to $x(t)$ since the GOBF approach cannot deal with process noise. Moreover, the output noise variance was lowered from \mbox{$\lambda_e = 0.1$} to \mbox{$\lambda_e = 0.01$} as the large output noise would otherwise dominate so much that no sensible conclusions could be made. 

One thousand Monte Carlo simulations are performed, with each time an estimation and a validation data set of \mbox{$N = 1000$} data points each. In case of the approximative PEM method~\citep{Hagenblad2008}, the true model structure is assumed to be known, and the true system belongs to the considered model set. In case of the proposed GOBF approach, the order of the linear dynamics is assumed to be known. A model with \mbox{$n_{rep} = 0$} and one with \mbox{$n_{rep} = 1$} is estimated. The local polynomial method~\citep{Pintelon2010} is used to estimate the BLA. The nonlinearity is approximated via a multivariate polynomial of degree $3$, in order to capture both even and odd nonlinearities. Note that in this case, the true system is not in the model set.

\begin{figure}
	\centering
	\includegraphics[width=0.4\textwidth]{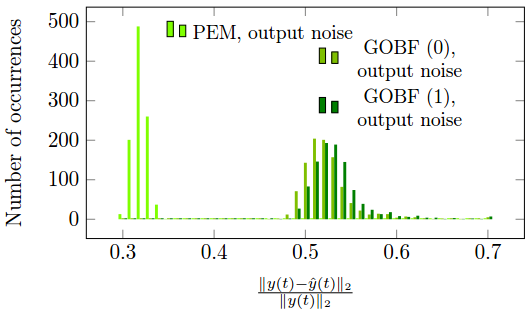}
	\caption{Distribution of $\frac{\left\Vert y(t) - \hat{y}(t) \right\Vert_2}{\left\Vert y(t) \right\Vert_2}$ on the validation data sets for the $1000$ Monte Carlo simulations. PEM indicates the results for the approximative prediction error method, while GOBF (0) and GOBF (1) indicate the results for the proposed GOBF approach with \mbox{$n_{rep} = 0$} and \mbox{$n_{rep} = 1$}.
\label{fig: histogram NRMSE output noise}}
\end{figure}

\begin{figure}
	\centering
	\includegraphics[width=0.4\textwidth]{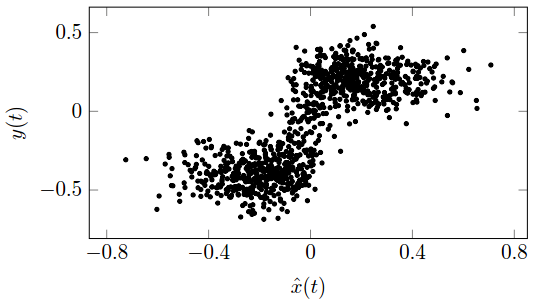}
	\caption{The scatter plot of $\hat{x}(t)$ and $y(t)$ reveals the shape of the saturation nonlinearity (GOBF approach with \mbox{$n_{rep} = 0$}, last Monte Carlo simulation).
	\label{fig: scatter plot}}
\end{figure}

The results in Fig.~\ref{fig: histogram NRMSE output noise} show that the approximative PEM method performs significantly better than the GOBF approach. Note that the approximative PEM method used full prior knowledge of the model structure, while no prior knowledge on the nonlinearity was used in the GOBF approach. Still, it is able to find a decent approximation. A better approximation can be obtained by representing the nonlinearity with another basis function expansion that is more appropriate to the nonlinearity at hand. This, however, is out of the scope of this paper.
Finally, for a single-branch Wiener system, the shape of the output nonlinearity can be determined as follows. Motivated by Lemma~\ref{lemma: bound on delta_x} and~\eqref{eq: BLA}, an estimate of the coefficients $\alpha$ can be obtained as
\begin{equation}
	\hat{\alpha} = \argmin_{\alpha} \left\lvert y(t) - \sum_{l=0}^n \alpha_l x_l(t) \right\rvert^2
	\quad .
\end{equation}
The signal \mbox{$\hat{x}(t) = \sum_{l=0}^n \alpha_l x_l(t)$} is then, up to an unknown scale factor $c_{BLA}$, approximately equal to $x(t)$. The shape of the nonlinear function $f$ can then be determined from a scatter plot of $\hat{x}(t)$ and $y(t)$ (see Fig.~\ref{fig: scatter plot}).

\begin{figure}
	\centering
	\includegraphics[width=0.4\textwidth]{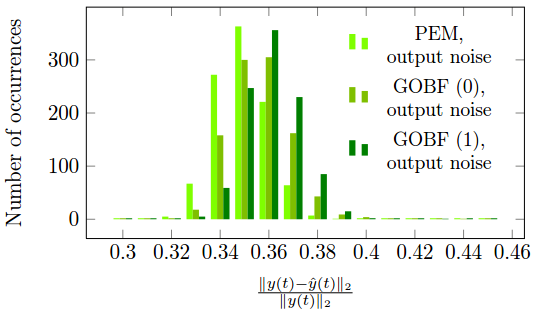}
	\caption{Distribution of $\frac{\left\Vert y(t) - \hat{y}(t) \right\Vert_2}{\left\Vert y(t) \right\Vert_2}$ on the validation data sets for the $1000$ Monte Carlo simulations (polynomial nonlinearity).
\label{fig: histogram NRMSE polynomial}}
\end{figure}

To make a fair comparison, the example is modified such that the system is in the model class for both of the considered approaches. The nonlinearity in~\eqref{eq: example 2} is changed to a third-degree polynomial that best approximates the saturation nonlinearity on all the estimation data sets.
The approximative PEM method now estimates a third-degree polynomial nonlinearity.
In this case, the GOBF approaches have a similar performance as the approximative PEM approach (see Fig.~\ref{fig: histogram NRMSE polynomial}).
Finally, in order to determine the number of repetitions $n_{rep}$, one can easily estimate several models for an increasing $n_{rep}$, and compare the simulation errors on a validation data set. Once the simulation error increases, the variance error outweighs the model error, and one should select less repetitions. Here, the normalized rms error for \mbox{$n_{rep} = 0$} is lower than the normalized rms error for \mbox{$n_{rep} = 1$} in $850$ out of the $1000$ cases. In the remaining $150$ cases, one would select a model with \mbox{$n_{rep} \ge 1$}.

\section{Conclusion}
\label{sec: Conclusion}

An identification procedure for SISO Wiener systems with finite-order IIR dynamics and a polynomial nonlinearity was formulated and its asymptotic behavior was analyzed in an output-error framework. It is shown that the estimated output converges in probability to the true system output. Fast convergence rates can be obtained.

The identification procedure is mainly linear in the parameters. The proposed identification procedure is thus well suited to provide an initial guess for nonlinear optimization algorithms. The approach can be applied to parallel Wiener systems as well.

\begin{ack}
This work was supported by the ERC advanced grant SNLSID, under contract 320378.
\end{ack}

\bibliographystyle{model5-names}

\begin{thebibliography}{31}
\expandafter\ifx\csname natexlab\endcsname\relax\def\natexlab#1{#1}\fi
\providecommand{\bibinfo}[2]{#2}
\ifx\xfnm\relax \def\xfnm[#1]{\unskip,\space#1}\fi
\bibitem[{Billings \& Fakhouri(1982)}]{Billings1982}
\bibinfo{author}{Billings, S.~A.}, \& \bibinfo{author}{Fakhouri, S.~Y.}
  (\bibinfo{year}{1982}).
\newblock \bibinfo{title}{Identification of systems containing linear dynamic
  and static nonlinear elements}.
\newblock {\it \bibinfo{journal}{Automatica}\/},  {\it \bibinfo{volume}{18}\/},
  \bibinfo{pages}{15--26}.
\bibitem[{Bloemen et~al.(2001)Bloemen, Chou, van~den Boom, Verdult, Verhaegen
  \& Backx}]{Bloemen2001}
\bibinfo{author}{Bloemen, H. H.~J.}, \bibinfo{author}{Chou, C.~T.},
  \bibinfo{author}{van~den Boom, T. J.~J.}, \bibinfo{author}{Verdult, V.},
  \bibinfo{author}{Verhaegen, M.}, \& \bibinfo{author}{Backx, T.~C.}
  (\bibinfo{year}{2001}).
\newblock \bibinfo{title}{{W}iener model identification and predictive control
  for dual composition control of a distillation column}.
\newblock {\it \bibinfo{journal}{Journal of Process Control}\/},  {\it
  \bibinfo{volume}{11}\/}, \bibinfo{pages}{601--620}.
\bibitem[{Boyd \& Chua(1985)}]{Boyd1985}
\bibinfo{author}{Boyd, S.}, \& \bibinfo{author}{Chua, L.~O.}
  (\bibinfo{year}{1985}).
\newblock \bibinfo{title}{Fading memory and the problem of approximating
  nonlinear operators with {Volterra} series}.
\newblock {\it \bibinfo{journal}{IEEE Transactions on Circuits and Systems}\/},
   {\it \bibinfo{volume}{32}\/}, \bibinfo{pages}{1150--1161}.
\bibitem[{Bussgang(1952)}]{Bussgang1952}
\bibinfo{author}{Bussgang, J.~J.} (\bibinfo{year}{1952}).
\newblock {\it \bibinfo{title}{Crosscorrelation functions of
  amplitude-distorted Gaussian signals}\/}.
\newblock \bibinfo{type}{Technical Report} \bibinfo{number}{216.} {MIT}
  research laboratory of electronics.
\bibitem[{Chua \& Ng(1979)}]{Chua1979}
\bibinfo{author}{Chua, L.~O.}, \& \bibinfo{author}{Ng, C.-Y.}
  (\bibinfo{year}{1979}).
\newblock \bibinfo{title}{Frequency domain analysis of nonlinear systems:
  General theory}.
\newblock {\it \bibinfo{journal}{IEE Journal on Electronic Circuits and
  Systems}\/},  {\it \bibinfo{volume}{3}\/}, \bibinfo{pages}{165--185}.
\bibitem[{Giri \& Bai(2010)}]{Giri2010}
\bibinfo{editor}{Giri, F.}, \& \bibinfo{editor}{Bai, E.-W.} (Eds.)
  (\bibinfo{year}{2010}).
\newblock {\it \bibinfo{title}{Block-oriented Nonlinear System
  Identification}\/}.
\newblock (\bibinfo{edition}{1st} ed.).
\newblock \bibinfo{publisher}{Springer}.
\bibitem[{Giri et~al.(2013)Giri, Rochdi, Radouane, Brouri \& Chaoui}]{Giri2013}
\bibinfo{author}{Giri, F.}, \bibinfo{author}{Rochdi, Y.},
  \bibinfo{author}{Radouane, A.}, \bibinfo{author}{Brouri, A.}, \&
  \bibinfo{author}{Chaoui, F.-Z.} (\bibinfo{year}{2013}).
\newblock \bibinfo{title}{Frequency identification of nonparametric {Wiener}
  systems containing backlash nonlinearities}.
\newblock {\it \bibinfo{journal}{Automatica}\/},  {\it \bibinfo{volume}{49}\/},
  \bibinfo{pages}{124--137}.
\bibitem[{Greblicki(1994)}]{Greblicki1994}
\bibinfo{author}{Greblicki, W.} (\bibinfo{year}{1994}).
\newblock \bibinfo{title}{Nonparametric identification of {Wiener} systems by
  orthogonal series}.
\newblock {\it \bibinfo{journal}{IEEE Transactions on Automatic Control}\/},
  {\it \bibinfo{volume}{39}\/}, \bibinfo{pages}{2077--2086}.
\bibitem[{Guillaume et~al.(1989)Guillaume, Schoukens \&
  Pintelon}]{Guillaume1989}
\bibinfo{author}{Guillaume, P.}, \bibinfo{author}{Schoukens, J.}, \&
  \bibinfo{author}{Pintelon, R.} (\bibinfo{year}{1989}).
\newblock \bibinfo{title}{Sensitivity of roots to errors in the coefficient of
  polynomials obtained by frequency-domain estimation methods}.
\newblock {\it \bibinfo{journal}{IEEE Transactions on Instrumentation and
  Measurement}\/},  {\it \bibinfo{volume}{38}\/}, \bibinfo{pages}{1050--1056}.
\bibitem[{Hagenblad et~al.(2008)Hagenblad, Ljung \& Wills}]{Hagenblad2008}
\bibinfo{author}{Hagenblad, A.}, \bibinfo{author}{Ljung, L.}, \&
  \bibinfo{author}{Wills, A.} (\bibinfo{year}{2008}).
\newblock \bibinfo{title}{Maximum likelihood identification of {W}iener
  models}.
\newblock {\it \bibinfo{journal}{Automatica}\/},  {\it \bibinfo{volume}{44}\/},
  \bibinfo{pages}{2697--2705}.
\bibitem[{Heuberger et~al.(1995)Heuberger, Van~den Hof \&
  Bosgra}]{Heuberger1995}
\bibinfo{author}{Heuberger, P. S.~C.}, \bibinfo{author}{Van~den Hof, P. M.~J.},
  \& \bibinfo{author}{Bosgra, O.~H.} (\bibinfo{year}{1995}).
\newblock \bibinfo{title}{A generalized orthonormal basis for linear dynamical
  systems}.
\newblock {\it \bibinfo{journal}{IEEE Transactions on Automatic Control}\/},
  {\it \bibinfo{volume}{40}\/}, \bibinfo{pages}{451--465}.
\bibitem[{Heuberger et~al.(2005)Heuberger, Van~den Hof \&
  Wahlberg}]{Heuberger2005}
\bibinfo{editor}{Heuberger, P. S.~C.}, \bibinfo{editor}{Van~den Hof, P. M.~J.},
  \& \bibinfo{editor}{Wahlberg, B.} (Eds.) (\bibinfo{year}{2005}).
\newblock {\it \bibinfo{title}{Modelling and Identification with Rational
  Orthogonal Basis Functions}\/}.
\newblock \bibinfo{address}{London}: \bibinfo{publisher}{Springer}.
\bibitem[{Hunter \& Korenberg(1986)}]{Hunter1986}
\bibinfo{author}{Hunter, I.~W.}, \& \bibinfo{author}{Korenberg, M.~J.}
  (\bibinfo{year}{1986}).
\newblock \bibinfo{title}{The identification of nonlinear biological systems:
  {W}iener and {H}ammerstein cascade models}.
\newblock {\it \bibinfo{journal}{Biological Cybernetics}\/},  {\it
  \bibinfo{volume}{55}\/}, \bibinfo{pages}{135--144}.
\bibitem[{Kalafatis et~al.(1995)Kalafatis, Arifin, Wang \&
  Cluett}]{Kalafatis1995}
\bibinfo{author}{Kalafatis, A.}, \bibinfo{author}{Arifin, N.},
  \bibinfo{author}{Wang, L.}, \& \bibinfo{author}{Cluett, W.~R.}
  (\bibinfo{year}{1995}).
\newblock \bibinfo{title}{A new approach to the identification of {pH}
  processes based on the {W}iener model}.
\newblock {\it \bibinfo{journal}{Chemical Engineering Science}\/},  {\it
  \bibinfo{volume}{50}\/}, \bibinfo{pages}{3693--3701}.
\bibitem[{Ljung(2013)}]{Ljung2013}
\bibinfo{author}{Ljung, L.} (\bibinfo{year}{2013}).
\newblock {\it \bibinfo{title}{System Identification Toolbox - User's Guide}\/}
  (\bibinfo{edition}{8th} ed.).
\bibitem[{Pelckmans(2011)}]{Pelckmans2011}
\bibinfo{author}{Pelckmans, K.} (\bibinfo{year}{2011}).
\newblock \bibinfo{title}{{MINLIP} for the identification of monotone {Wiener}
  systems}.
\newblock {\it \bibinfo{journal}{Automatica}\/},  {\it \bibinfo{volume}{47}\/},
  \bibinfo{pages}{2298--2305}.
\bibitem[{Pintelon et~al.(1994)Pintelon, Guillaume, Rolain, Schoukens \&
  Van~hamme}]{Pintelon1994}
\bibinfo{author}{Pintelon, R.}, \bibinfo{author}{Guillaume, P.},
  \bibinfo{author}{Rolain, Y.}, \bibinfo{author}{Schoukens, J.}, \&
  \bibinfo{author}{Van~hamme, H.} (\bibinfo{year}{1994}).
\newblock \bibinfo{title}{Parametric identification of transfer functions in
  the frequency domain -- a survey}.
\newblock {\it \bibinfo{journal}{IEEE Transactions on Automatic Control}\/},
  {\it \bibinfo{volume}{39}\/}, \bibinfo{pages}{2245--2260}.
\bibitem[{Pintelon \& Schoukens(2012)}]{Pintelon2012}
\bibinfo{author}{Pintelon, R.}, \& \bibinfo{author}{Schoukens, J.}
  (\bibinfo{year}{2012}).
\newblock {\it \bibinfo{title}{System Identification: A Frequency Domain
  Approach}\/}.
\newblock (\bibinfo{edition}{2nd} ed.).
\newblock \bibinfo{publisher}{Wiley-IEEE Press}.
\bibitem[{Pintelon et~al.(2010)Pintelon, Schoukens, Vandersteen \&
  Barb\'e}]{Pintelon2010}
\bibinfo{author}{Pintelon, R.}, \bibinfo{author}{Schoukens, J.},
  \bibinfo{author}{Vandersteen, G.}, \& \bibinfo{author}{Barb\'e, K.}
  (\bibinfo{year}{2010}).
\newblock \bibinfo{title}{Estimation of nonparametric noise and {FRF} models
  for multivariable systems-part~{I}: Theory}.
\newblock {\it \bibinfo{journal}{Mechanical Systems and Signal Processing}\/},
  {\it \bibinfo{volume}{24}\/}, \bibinfo{pages}{573--595}.
\bibitem[{da~Rosa et~al.(2007)da~Rosa, Campello \& Amaral}]{daRosa2007}
\bibinfo{author}{da~Rosa, A.}, \bibinfo{author}{Campello, R. J. G.~B.}, \&
  \bibinfo{author}{Amaral, W.~C.} (\bibinfo{year}{2007}).
\newblock \bibinfo{title}{Choice of free parameters in expansions of
  discrete-time {Volterra} models using {Kautz} functions}.
\newblock {\it \bibinfo{journal}{Automatica}\/},  {\it \bibinfo{volume}{43}\/},
  \bibinfo{pages}{1084--1091}.
\bibitem[{Schetzen(2006)}]{Schetzen2006}
\bibinfo{author}{Schetzen, M.} (\bibinfo{year}{2006}).
\newblock {\it \bibinfo{title}{The {Volterra} \& {Wiener} Theories of Nonlinear
  Systems}\/}.
\newblock \bibinfo{address}{Malabar, Florida}: \bibinfo{publisher}{Krieger
  Publishing Company}.
\bibitem[{Schoukens et~al.(1998)Schoukens, Dobrowiecki \&
  Pintelon}]{Schoukens1998}
\bibinfo{author}{Schoukens, J.}, \bibinfo{author}{Dobrowiecki, T.}, \&
  \bibinfo{author}{Pintelon, R.} (\bibinfo{year}{1998}).
\newblock \bibinfo{title}{Parametric and nonparametric identification of linear
  systems in the presence of nonlinear distortions - a frequency domain
  approach}.
\newblock {\it \bibinfo{journal}{IEEE Transactions on Automatic Control}\/},
  {\it \bibinfo{volume}{43}\/}, \bibinfo{pages}{176--190}.
\bibitem[{Schoukens et~al.(2009)Schoukens, Lataire, Pintelon, Vandersteen \&
  Dobrowiecki}]{Schoukens2009a}
\bibinfo{author}{Schoukens, J.}, \bibinfo{author}{Lataire, J.},
  \bibinfo{author}{Pintelon, R.}, \bibinfo{author}{Vandersteen, G.}, \&
  \bibinfo{author}{Dobrowiecki, T.} (\bibinfo{year}{2009}).
\newblock \bibinfo{title}{Robustness issues of the best linear approximation of
  a nonlinear system}.
\newblock {\it \bibinfo{journal}{IEEE Transactions on Instrumentation and
  Measurement}\/},  {\it \bibinfo{volume}{58}\/}, \bibinfo{pages}{1737--1745}.
\bibitem[{Schoukens \& Rolain(2012)}]{Schoukens2012}
\bibinfo{author}{Schoukens, M.}, \& \bibinfo{author}{Rolain, Y.}
  (\bibinfo{year}{2012}).
\newblock \bibinfo{title}{Parametric identification of parallel {Wiener}
  systems}.
\newblock {\it \bibinfo{journal}{IEEE Transactions on Instrumentation and
  Measurement}\/},  {\it \bibinfo{volume}{61}\/}, \bibinfo{pages}{2825--2832}.
\bibitem[{Tiels \& Schoukens(2011)}]{Tiels2011}
\bibinfo{author}{Tiels, K.}, \& \bibinfo{author}{Schoukens, J.}
  (\bibinfo{year}{2011}).
\newblock \bibinfo{title}{Identifying a {Wiener} system using a variant of the
  {Wiener G-Functionals}}.
\newblock In {\it \bibinfo{booktitle}{{$50^{th}$} {IEEE} {Conference} on
  {Decision} and {Control} and {European} {Control} {Conference} ({CDC-ECC11}),
  Orlando, FL, USA}\/}.
\bibitem[{van~der Vaart(1998)}]{vanderVaart1998}
\bibinfo{author}{van~der Vaart, A.~W.} (\bibinfo{year}{1998}).
\newblock {\it \bibinfo{title}{Asymptotic Statistics}\/}.
\newblock \bibinfo{publisher}{Cambridge University Press}.
\bibitem[{{Van den Hof} et~al.(1995){Van den Hof}, Heuberger \&
  Bokor}]{VandenHof1995}
\bibinfo{author}{{Van den Hof}, P. M.~J.}, \bibinfo{author}{Heuberger, P.
  S.~C.}, \& \bibinfo{author}{Bokor, J.} (\bibinfo{year}{1995}).
\newblock \bibinfo{title}{System identification with generalized orthonormal
  basis functions}.
\newblock {\it \bibinfo{journal}{Automatica}\/},  {\it \bibinfo{volume}{31}\/},
  \bibinfo{pages}{1821--1834}.
\bibitem[{de~Vries \& Van~den Hof(1998)}]{deVries1998}
\bibinfo{author}{de~Vries, D.~K.}, \& \bibinfo{author}{Van~den Hof, P. M.~J.}
  (\bibinfo{year}{1998}).
\newblock \bibinfo{title}{Frequency domain identification with generalized
  orthonormal basis functions}.
\newblock {\it \bibinfo{journal}{IEEE Transactions on Automatic Control}\/},
  {\it \bibinfo{volume}{43}\/}, \bibinfo{pages}{656--669}.
\bibitem[{Westwick \& Verhaegen(1996)}]{Westwick1996}
\bibinfo{author}{Westwick, D.}, \& \bibinfo{author}{Verhaegen, M.}
  (\bibinfo{year}{1996}).
\newblock \bibinfo{title}{Identifying {MIMO} {Wiener} systems using subspace
  model identification methods}.
\newblock {\it \bibinfo{journal}{Signal Processing}\/},  {\it
  \bibinfo{volume}{52}\/}, \bibinfo{pages}{235--258}.
\bibitem[{Wiener(1958)}]{Wiener1958}
\bibinfo{author}{Wiener, N.} (\bibinfo{year}{1958}).
\newblock {\it \bibinfo{title}{Nonlinear problems in random theory}\/}.
\newblock \bibinfo{publisher}{Wiley}.
\bibitem[{Wills et~al.(2013)Wills, Sch\"on, Ljung \& Ninness}]{Wills2013}
\bibinfo{author}{Wills, A.}, \bibinfo{author}{Sch\"on, T.~B.},
  \bibinfo{author}{Ljung, L.}, \& \bibinfo{author}{Ninness, B.}
  (\bibinfo{year}{2013}).
\newblock \bibinfo{title}{Identification of {Hammerstein-Wiener} models}.
\newblock {\it \bibinfo{journal}{Automatica}\/},  {\it \bibinfo{volume}{49}\/},
  \bibinfo{pages}{70--81}.

\end{thebibliography}

\appendix
\section{Orthonormal basis}
\label{app: Orthonormal basis}
The orthogonality of $F_0(z)$ with respect to the set $\{F_l(z)\}$ (\mbox{$l = 1, 2, \ldots$}) can be shown either by working out the inner products $\langle F_l(z) , F_0(z) \rangle$ (see below) or by choosing a pole structure \mbox{$\{0, \xi_1, \xi_2, \ldots\}$} in the shifted basis functions \mbox{$F_l^{S}(z) := z F_l(z)$}.

\begin{pf}
Consider the basis functions $F_{l}^{B}(z)$, given by
\mbox{$F_{j + (k - 1) n_{\xi}}^{B}(z) = \frac{1}{(z - \xi_j)^{k}}$} for \mbox{$j = 1, \ldots, n_{\xi}$} and \mbox{$k = 1, 2, \ldots$}.
We now prove that they are orthogonal to \mbox{$F_0(z) = 1$}, by showing that the inner product
\begin{equation}
	\begin{aligned}
	&\langle F_{j + (k - 1) n_{\xi}}^{B}(z) , F_{0}(z) \rangle \\ &\qquad = 
	\frac{1}{2 \pi i} \oint_{\mathbb{T}}
		F_{j + (k - 1) n_{\xi}}^{B}(z) F_{0}^{*}\left(\frac{1}{z^{*}}\right) \frac{\mathrm{d}z}{z}
	\quad ,
	\end{aligned}
\end{equation}
is equal to zero. $\mathbb{T}$ denotes the unit circle.
\begin{enumerate}
	\item $\xi_j \ne 0$\\
		\begin{equation} \begin{aligned}
			&\langle F_{j + (k - 1) n_{\xi}}^{B}(z) , F_{0}(z) \rangle
				\\ &\qquad =  \frac{1}{2 \pi i} \oint_{\mathbb{T}} \frac{1}{z(z - \xi_j)^{k}} \mathrm{d}z 		\\
				\\ &\qquad = \frac{1}{(k - 1)!} \lim_{z \rightarrow \xi_j} 
						\frac{\mathrm{d}^{k - 1}}{\mathrm{d}z^{k - 1}} \left(\frac{1}{z}\right)
					+ \lim_{z \rightarrow 0} \frac{1}{(z - \xi_j)^{k}}					\\
				\\ &\qquad = \frac{1}{(k - 1)!} (-1)^{k - 1} (k - 1)! \, \xi_{j}^{- k} + (- 1)^{k} \xi_{j}^{- k} 		\\
				\\ &\qquad = \xi_{j}^{- k} \left( (- 1)^{k - 1} + (- 1)^{k} \right) 					\\
				\\ &\qquad = 0
		\end{aligned} \end{equation}
	\item $\xi_j = 0$\\
		\begin{equation} \begin{aligned}
			\langle F_{j + (k - 1) n_{\xi}}^{B}(z) , F_{0}(z) \rangle
				& =  \frac{1}{2 \pi i} \oint_{\mathbb{T}} \frac{1}{z^{k + 1}} \mathrm{d}z 		\\
				& = \frac{1}{k!} \lim_{z \rightarrow 0}
						\left. \left. \frac{\mathrm{d}^k}{\mathrm{d}z^{k}} \right( 1 \right) 	\\
				& = 0
		\end{aligned} \end{equation}
\end{enumerate}
The OBFs $F_l(z)$ are linear combinations of the basis functions $F_{l}^{B}(z)$~\citep{Heuberger2005} and are thus orthogonal to \mbox{$F_{0}(z) = 1$}. Since the norm of $F_{0}(z)$ is equal to one, the set $\{F_l(z)\}$ (\mbox{$l = 0, 1, \ldots$}) is a set of OBFs. \qed
\end{pf}

\section{Volterra series}
\label{app: Volterra series}
A Volterra series generalizes the impulse response of an LTI system to a nonlinear time-invariant system via multidimensional impulse responses. The input-output relation of a Volterra series is split in different contributions of increasing degree of nonlinearity~\citep{Schetzen2006}
\begin{subequations}
\begin{equation}
	y(t) = \sum_{p = 1}^{\infty} y_p(t)
	\quad ,
\end{equation}
with
\begin{multline}
	y_p(t) = \int_{- \infty}^{+ \infty} \cdots \int_{- \infty}^{+ \infty} h_p(\tau_1, \ldots, \tau_p)\\
			u(t - \tau_1) \cdots u(t - \tau_p) \mathrm{d}\tau_1 \cdots \mathrm{d}\tau_p
	\quad ,
\end{multline}
\end{subequations}
where $h_p$ is the Volterra kernel of degree $p$.

For periodic excitations, the output Fourier coefficient $Y_p(k)$ at frequency $\frac{k f_{max}}{N_F}$ is~\citep{Chua1979}
\begin{multline}
	Y_p(k) = \sum_{k_1 = - N_F}^{N_F} \sum_{k_2 = - N_F}^{N_F} \cdots \sum_{k_{p - 1} = - N_F}^{N_F}\\
			H_p(L_k, k_1, k_2, \ldots, k_{p - 1})\\ U(k_1) U(k_2) \cdots U(k_{p - 1}) U(L_k)
	\quad ,
\end{multline}
with $L_k = k - \sum_{i = 1}^{p - 1} k_i$.
Here, $H_p$ is the symmetrized frequency domain representation of the Volterra kernel of degree $p$.
The Volterra series is uniformly bounded if~\citep{Schoukens1998}
\begin{equation}
	\sum_{p = 1}^{\infty} M_{H_p} {M_U}^p \le C_1 < \infty
	\quad ,
\end{equation}
with \mbox{$M_{H_p} = \max \left| H_p \right|$}, and $M_U$ as in Definition~\ref{def: excitation signal class}.

\end{document}